
\def\a{{\alpha}}
\def\b{{\beta}}
\def\g{{\gamma}}
\def\k{{\kappa}}

\def\bz{{\bf z}}

\def\bp{{\bf p}}
\def\bm{{\bf m}}
\def\bn{{\bf n}}
\def\bo{{\bf 1}}
\def\baz{{\bar z}}
\def\mnt{M_{\bf n}(t)}
\def\mn{M_{\bf n}}
\def\cnt{{\cal N}(t)}

\def\pmt{{P(\bm,t)}}
\def\pbm{{P(\bm)}}
\def\pmp{{P(m,p)}}
\def\gm{{g(\bm)}}
\def\fzt{{F(\bz,t)}}
\def\fz{{F(\bz)}}
\def\rbm{{R(\bm)}}
\def\rz{{R(\bz)}}

\def\summ{{\sum_{\{m_i\}}}}
\def\sump{{\sum_{\{m_i'\}}}}

\def\vav{{\langle v\rangle}}
\def\mav{{\langle m\rangle}}
\def\mavr{{\langle m(r)\rangle}}

\def\pav{{\langle |p|\rangle}}

\def\eg{{\it e.\ g.}}\def\ie{{\it i.\ e.}}
\def\ea{{\it et al.}}

\newcount\refnum\refnum=0  
\def\refi{\smallskip\global\advance\refnum by 1\item{\the\refnum.}}

\newcount\eqnum \eqnum=0  
\def\eqnoi{\global\advance\eqnum by 1\eqno(\the\eqnum)}
\def\eqnai{\global\advance\eqnum by 1\eqno(\the\eqnum {\rm a})}
\def\back#1{{\advance\eqnum by-#1 Eq.~(\the\eqnum)}}
\def\last{Eq.~(\the\eqnum)~}                   

\def\p2d#1#2{{\partial^2 #1\over\partial #2^2}} 
\def\td#1#2{{d #1\over d #2}}      
\def\t2d#1#2{{d^2 #1\over d #2^2}} 

\def\acp #1 #2 #3 {{\sl Adv.\ Chem.\ Phys.} {\bf #1}, #2 (#3)}
\def\cp #1 #2 #3 {{\sl Chem.\ Phys.} {\bf #1}, #2 (#3)}
\def\eul #1 #2 #3 {{\sl Europhys.\ Lett.} {\bf #1}, #2 (#3)}
\def\jcp #1 #2 #3 {{\sl J.\ Chem.\ Phys.} {\bf #1}, #2 (#3)}
\def\jdep #1 #2 #3 {{\sl J.\ de Physique I} {\bf #1}, #2 (#3)}
\def\jdepl #1 #2 #3 {{\sl J. de Physique Lett.} {\bf #1}, #2 (#3)}
\def\jetp #1 #2 #3 {{\sl Sov.\ Phys.\ JETP} {\bf #1}, #2 (#3)}
\def\jetpl #1 #2 #3 {{\sl Sov. Phys.\ JETP Letters} {\bf #1}, #2 (#3)}
\def\jmp #1 #2 #3 {{\sl J. Math. Phys.} {\bf #1}, #2 (#3)}
\def\jpa #1 #2 #3 {{\sl J. Phys.\ A} {\bf #1}, #2 (#3)}
\def\jpchem #1 #2 #3 {{\sl J. Phys.\ Chem.} {\bf #1}, #2 (#3)}
\def\jpsj #1 #2 #3 {{\sl J. Phys.\ Soc. Jpn.} {\bf #1}, #2 (#3)}
\def\jsp #1 #2 #3 {{\sl J. Stat.\ Phys.} {\bf #1}, #2 (#3)}
\def\nat #1 #2 #3 {{\sl Nature} {\bf #1}, #2 (#3)}
\def\pA #1 #2 #3 {{\sl Physica A} {\bf #1}, #2 (#3)}
\def\pD #1 #2 #3 {{\sl Physica D} {\bf #1}, #2 (#3)}
\def\pla #1 #2 #3 {{\sl Phys.\ Lett. A} {\bf #1}, #2 (#3)}
\def\plb #1 #2 #3 {{\sl Phys.\ Lett. B} {\bf #1}, #2 (#3)}
\def\pr #1 #2 #3 {{\sl Phys.\ Rev.} {\bf #1}, #2 (#3)}
\def\pra #1 #2 #3 {{\sl Phys.\ Rev.\ A} {\bf #1}, #2 (#3)}
\def\prb #1 #2 #3 {{\sl Phys.\ Rev.\ B} {\bf #1}, #2 (#3)}
\def\pre #1 #2 #3 {{\sl Phys.\ Rev.\ E} {\bf #1}, #2 (#3)}
\def\prl #1 #2 #3 {{\sl Phys.\ Rev.\ Lett.} {\bf #1}, #2 (#3)}
\def\prslA #1 #2 #3 {{\sl Proc.\ R.\ Soc.\ London, Ser.\ A} {\bf #1}, #2 (#3)}
\def\rmp #1 #2 #3 {{\sl Rev.\ Mod.\ Phys.} {\bf #1}, #2 (#3)}
\def\sci #1 #2 #3 {{\sl Science} {\bf #1}, #2 (#3)}
\def\sciam #1 #2 #3 {{\sl Scientific American} {\bf #1}, #2 (#3)}
\def\zpb #1 #2 #3 {{\sl Z. Phys.\ B} {\bf #1}, #2 (#3)}
\def\zpc #1 #2 #3 {{\sl Z. Phys.\ Chem.} {\bf #1}, #2 (#3)}

\overfullrule=0pt
\magnification 1200
\baselineskip=18 true bp

\centerline{\bf Aggregation with Multiple Conservation Laws}
\bigskip\bigskip
\centerline{\bf P.~L.~Krapivsky$^1$ and E.~Ben-Naim$^2$}
\medskip
\centerline{$^1$Center for Polymer Studies and Department of Physics}
\centerline{Boston University, Boston, MA 02215, USA}
\medskip
\centerline{$^2$The James Franck Institute}
\centerline{The University of Chicago, Chicago, IL 60637, USA}
\bigskip\bigskip
\centerline{\bf Abstract}
{\noindent Aggregation processes with an arbitrary
number of conserved quantities are investigated. On the mean-field
level, an exact solution for the size distribution is obtained.  The
asymptotic form of this solution exhibits nontrivial
``double'' scaling. While processes with one conserved quantity are
governed by a single scale, processes with multiple conservation laws
exhibit an additional diffusion-like scale. The theory is applied
to ballistic aggregation with mass and momentum conserving collisions
and to diffusive aggregation with multiple species.
\bigskip}
\vskip 0.6in
{\noindent P. A. C. S. Numbers: 02.50.-r, 05.40.+j, 82.20.-w}

\vfil\eject

\centerline{\bf I. Introduction}\medskip

Irreversible aggregation processes underly many natural
phenomena including, \eg, polymerization [1], gelation [2],
island growth [3-4] and aerosols [5].
The classical rate theory of Smoluchowski describes the kinetics
of such processes [6-9]. Recently, scaling [10-16] and
exact [17-23] theoretical studies showed that spatial
correlations play a crucial role in low dimensions.
While the above examples are diffusive driven,
there are physical situations such as formation of large-scale
structure of the universe [24] and clustering in
traffic flows [25], where the aggregates move ballistically.
So far, theories of ballistic aggregation [26] have been
restricted to scaling arguments [26-30].

In the ballistic aggregation process both the mass and the momentum
are conserved. In polymerization processes involving copolymers, each
monomer species mass is a conserved quantity. Hence, we study
aggregation processes with multiple conservation laws. The simplest
example for such a system is aggregation with $k$ distinct species.
Both the multivariate distribution and the single variable
distributions are of interest. We present exact solutions to the
time dependent and the steady state mean-field rate equations.
Although they are straightforward generalizations to the well known
solutions they exhibit interesting behaviors.
An asymptotic analysis shows that fluctuations associated with
a single conserved quantity are Gaussian in nature.
As a result, an additional ``diffusive'' size scale emerges.

We apply the above theory to ballistic aggregation as well as
diffusive aggregation. In the case of ballistic aggregation, we use an
approximate collision rate to obtain a solution to the Boltzmann
equation in arbitrary dimension. While this approach agrees with the
scaling arguments, it suggests that for a given mass, the momentum
distribution is Gaussian.  We compare these predictions with one and
two dimensional simulations.  Furthermore, we consider steady state
properties of the aggregation process by introducing input of
particles.  For homogeneous input, a novel time scale describing the
density relaxation is found.  In the case of a localized input,
clustering occurs only for $d\le 2$.  Additionally, we apply the
theory to two-species aggregation with diffusing particles. Using the
density dependent reaction rate, we obtain the leading scaling
behavior of the two relevant mass scales.

This paper is organized as follows. In section II, we present exact
solutions of the rate equation theory. We investigate time dependent
as well as steady state properties of the process.  We then apply the
theory to ballistic aggregation with momentum conserving collisions
(section III) and to diffusive driven aggregation (section IV). We conclude
with a discussion and suggestions for further research in section V.

\vfill\eject
\centerline{\bf II. Theory}\medskip

Following the above discussion, there are aggregation processes
where several physical quantities are conserved.  In such
processes, it is natural to label the aggregates by a ``mass'' vector
$\bm\equiv(m_1,\cdots,m_k)$, where every component represents a
conservation law.  Let us denote the probability distribution
function for particles of mass $\bm$ at time $t$ by $\pmt$.
Mean field theory of the binary reaction process assumes that
reaction proceeds with a rate proportional to the product of the
reactants densities. Thus the mean field approximation  neglects
spatial correlations and therefore typically holds in dimensions
larger than some critical dimension $d_c$ [10, 14-16].
The rate equations [6] are written as
$$
\td \pmt t=
\sump P({\bf m'},t)P(\bm-{\bf m'},t)- 2\pmt\sump P({\bf m'},t),\eqnoi
$$
where the sum is carried over all $k$ variables $m'_i$, $1\le
i \le k$.  The loss term prefactor reflects the fact that two
particles are lost in each collision.  One can verify that the rate
equations conserve each mass separately, \ie,
$\sum m_i\pmt\equiv\sum m_i P_0(\bm)$, where $P_0(\bm)$
is the initial distribution. In writing \last we have implicitly
assumed that the rate $K({\bf m},{\bf n})$ at which the reaction
$({\bf m})+({\bf n})\to ({\bf m+n})$ proceeds does not depend on
masses of the reactants, $K({\bf m},{\bf n})={\rm const}$.
This constant can be set to unity without loss of generality.
For an arbitrary reaction rate kernel $K({\bf m},{\bf n})$, \last is
easily generalized so that, \eg, the gain term is replaced by
$\sum K({\bf m'},{\bf m-m'})P({\bf m'},t)P(\bm-{\bf m'},t)$.

To solve \last we introduce the generating function, $\fzt$,
defined by
$$
\fzt=\summ \bz^{\bm}\pmt.\eqnoi
$$
In the above equation we have used the shorthand notations
$\bz\equiv(z_1,\cdots,z_k)$ and $\bz^{\bm}\equiv z_1^{m_1}\cdots z_k^{m_k}$.
Moments of the distribution function are
readily obtained by evaluating the generating function and its
derivatives in the vicinity of the point $\bz=\bo\equiv
(1,\ldots,1)$. For example, the total cluster density is given
by $N(t)=F({\bf 1},t)$. The equation describing the temporal
evolution of the generating function  $\fzt$ can be evaluated by
a proper summation of the rate equation. This yields
$$
\td F t=F^2-2FN. \eqnoi
$$
As a preliminary step, we evaluate the time dependence of the
density. The corresponding rate equation is obtained by evaluating
\last\ at $\bz=\bo$,
$$
\td N t=-N^2. \eqnoi
$$
Without loss of generality we set the initial density to unity and
therefore we have $N(t)=1/(1+t)$. We also note that by subtracting
\last\ from \back1, a simple differential equation for the quantity
$F-N$, emerges, $d(F-N)/dt=(F-N)^2$. This equation is readily solved
to find
$$
\fzt={F_0(\bz)\over(1+t)\left(1+t-tF_0(\bz)\right)},
\eqnoi
$$
where the notation $F_0(\bz)\equiv F(\bz,t=0)$ has been used.
\last represents the general solution to \back4. Indeed, it
is a simple generalization of the well known solution [6] for
the single mass aggregation process.

We consider the simplest multivariate case,
$$
P_0(\bm)=k^{-1}\sum_{i=1}^k \delta(m_i-1)\prod_{j\ne i}\delta(m_j).\eqnoi
$$
These initial conditions imply $F_0\equiv\baz=(z_1+\cdots+z_k)/k$ and
from \back1\ we have
$$
\fzt={\baz\over (1+t)\left(1+t-t\baz\right)}.\eqnoi
$$
A possible application of these initial conditions is to aggregation
processes involving $k$ distinct species with equal initial densities.
Although this symmetric situation appears to be too simple at first, it
contains the necessary ingredients for exploring the long time kinetics.

Before we investigate the multivariate distribution,
let us first study properties of the following distribution function
$$
P(m_+,t)=\summ \pmt \delta \left(m_+-(m_1+\cdots+m_k)\right).\eqnoi
$$
This distribution corresponds to the sum variable
$m_+=m_1+\cdots+m_k$.  It is useful to introduce the generating
function $F(z,t)=\sum z^{m_+}P(m_+,t)$.  This generating function can
be obtained from $\fzt$ by replacing $\baz$ with $z$. Expansion of
$F(z,t)=z/(1+t)(1+t-tz)$ in powers of $z$ yields the sum distribution
$$
P(m_+,t)={t^{m_+-1}\over(1+t)^{m_+ +1}}.\eqnoi
$$
The variable $m_+$ ignores the ``identity'' of the different
conserved variables and thus, the problem reduces to ordinary aggregation.
In the long time limit, $m_+\sim t$, and the corresponding distribution
 is given by
$$
P(m_+,t)\simeq t^{-2}\exp(-m_+/t).\eqnoi
$$
Using the scaling variable, $M_+=m_+/t$, this distribution can be also
written in terms of a scaling function $P(m_+,t)\simeq t^{-2}\Psi(M_+)$,
with $\Psi(x)=\exp(-x)$.

The multivariate probability distribution function can be now found
by expanding  $\fzt$ and comparing with \back8,
$$
\pmt=P(m_+,t)g(\bm)\qquad{\rm with}\qquad \gm=
{k^{-m_+}(m_+)!\over m_1!\cdots m_k!}.\eqnoi
$$
Note that $\summ\gm=1$. The above expression is the explicit solution
to the mean field equations. However, its long time nature
is of particular interest and we proceed with an asymptotic analysis.

To study the asymptotic properties of
$\pmt$, we concentrate on the case $k=2$ and then generalize the
results to arbitrary $k$. The time independent geometric factor reads,
\hbox{$g(m_1,m_2)=2^{-m_+}(m_1+m_2)!/m_1!m_2!\sim
m_+^{-1/2}\exp\big(-(m_1-m_2)^2/2m_+\big)$}. The limit of large
masses, $m_1,m_2\gg1$, is the relevant one since the masses grow
indefinitely. Using the limiting form of $P(m_+,t)$, we
arrive at the asymptotic form of the mass density,
$$
P(m_1,m_2,t)\sim t^{-2}m_+^{-1/2}\exp(-m_+/t)\exp(-m_-^2/2m_+),\eqnoi
$$
with $m_{\pm}=m_1\pm m_2$. Two mass scales govern the mass
distribution, $m_+\sim t$ and $|m_-|\sim \sqrt{m_+} \sim \sqrt{t}$.
Furthermore, introduction of the scaling variables,
$M_+=m_+/t$ and $M_-=m_-/\sqrt{t}$, enables us to write
the solution in a convenient scaling form
$$
P(m_1,m_2,t)\sim t^{-5/2} \Phi(M_+,M_-).\eqnoi
$$
The scaling function $\Phi$ depends on the scaling variables only,
$$
\Phi(x,y)=x^{-1/2}\exp(-x)\exp(-y^2/2x).\eqnoi
$$

Interestingly, each of the variables, $m_1$ and $m_2$, exhibit simple
scaling: $m_1,m_2\sim t$. Additionally, the moments $\mnt$ of the
distribution function, $\mnt=\summ \bm^{\bn}\pmt$, are described
asymptotically by linear exponents. By taking derivatives of the
generating function, one can show that $\mn(t)\sim t^{\alpha(\bf n)}$
with $\alpha({\bf n})=n_+-1$ .  This behavior agrees with ordinary
scaling and therefore one could naively expect single-size scaling
to hold.  However, from the complete form of the mass distribution we
learn that it is {\it impossible} to write the scaling solution in
terms of a simple scaling function, $P(m_1,m_2)\sim
t^{-2}\Phi(m_1/t,m_2/t)$.  Such a scaling solution would imply that
the process has only one intrinsic size scale. Instead, this problem
exhibits ``double-scaling'', as the process is governed by two
distinct size scales. The second scale $|m_-| \sim \sqrt{t}$ is
hidden, \eg, it does not clearly appear in the moments of the
distribution function.  The mass difference is reminiscent of a
diffusive scale for the following reason. The mass difference,
$m_-=m_1-m_2$, is also a conserved variable, since each of its
components is conserved.  Hence, for an aggregate of total mass $m_+$,
the mass difference is a sum of $m_+$ random variables. According to
the central limit theorem this variable is Gaussian and thus,
$m_-\sim\sqrt{m_+}$.

The above results can be generalized to arbitrary $k$.
The generalized mass difference variable is
$m_-^2=k^{-1}\sum_{i,j} (m_i-m_j)^2$ and the mass distribution
follows the scaling form
$$
\pmt\sim t^{-(k+3)/2}\Phi(M_+,M_-),\eqnoi
$$
with the scaling function
$$
\Phi(x,y)=x^{-(k-1)/2}\exp(-x)\exp(-y^2/2x).\eqnoi
$$

The above results can be also generalized to situations with asymmetric
initial conditions, \ie, the initial conditions are not invariant with
respect to a permutation of the mass variables, $\{m_i\}$. Denoting
by $M_i=\summ m_i P_0(\bm)$  the $i^{\rm th}$ conserved mass,
one can easily show that with the transformation
$m_i\to m_i/M_i$ and $\pbm\to M_1\cdots M_k\pbm$, the system reduces
to the symmetric case. For this transformation to be valid,
the process must be truly multivariate, $M_i\ne 0$. In this
case, the prefactor $\gm$ equals probabilities associated
with a biased random walk in a $k$-dimensional ``mass'' space.

Previous results have been established for a
constant rate kernel, $K({\bf m},{\bf n})={\rm const}$.
Although the most general situation is hardly tractable,
in the case of sum-variable dependent reaction rate,
$K({\bf m},{\bf n})\equiv K(m_+, n_+)$, the sum-variable distribution
function $P(m_+, t)$ satisfies a simpler single-variable Smoluchowski
equation. In this case, the complete multivariate distribution function
$P({\bf m}, t)$ is related to $P(m_+, t)$
by \back5. Therefore solvable variants of the single-variable
Smoluchowski equation provide exact solutions for multivariate
Smoluchowski equation with the corresponding reaction rates. For the usual
Smoluchowski equation, exact solutions have been found for three
types of reaction rates: $K(m,n)={\rm const}, ~K(m,n)=m+n, ~K(m,n)=mn$,
and for their linear combinations [6,5,31]. In particular,
for the sum-kernel $K({\bf m},{\bf n})={1\over 2}(m_+ + n_+)$,
subject to the monodisperse initial conditions of \back{10},
we again arrive at the exact solution of \back5 with
$$
P(m_+,t)=\exp\bigl[-t-m_+(1-e^{-t})\bigr]~
{\bigl[m_+(1-e^{-t})\bigr]^{m_+-1}\over (m_+)!}.
\eqnoi
$$
Similarly, one can find an exact solution for the product kernel.

We now investigate the steady state properties of the aggregation
process by introducing input of particles into the system.
 For simplicity, we consider  homogeneous input with rate $h$
and restrict ourselves to the constant rate kernel.
The governing equations are modified by adding an input term,
$$
\td \pmt t=\sump P({\bf m'},t)P(\bm-{\bf
m'},t)- 2\pmt\sump P({\bf m'},t)+h\rbm.\eqnoi
$$
The input function, $\rbm$, satisfies the normalization condition
$\summ\rbm=1$ since the total input $h\summ\rbm$ is equal to $h$.

The solution to the above equation parallels the solution to the time
dependent problem.  We denote the steady state distribution by $\pbm$
and the steady state generating function by $\fz$. This generating function
is obtained by eliminating the time derivative of \last,
$F^2-2FN+hR=0$, where $R\equiv\rz=\summ \bz^{\bm} \rbm$.
The normalization condition $R(\bo)=1$ is satisfied by the input function,
and consequently the solution to the generating function reads
$$
\fz=\sqrt{h}\left(1-\sqrt{1-\rz}\right).\eqnoi
$$
The total monomer density is given by $N=F(\bo)$ and thus,
$$
N(h)=\sqrt{h}.\eqnoi
$$

In analogy with the initial conditions of \back{14}, we  consider the
input $\rbm=P_0(\bm)$, which in turn implies $\rz=\baz$.  In this
case, the steady state generating function is simply
\hbox{$\fz=\sqrt{h}\big(1-\sqrt{1-\baz}\big)$}.
Expanding in powers of the variables $z_i$, and comparing with
the definition of the  generating function,
one finds the steady state distribution
$$
\pbm=P(m_+)\gm,\qquad{\rm with}\qquad
P(m_+)={(2m_+)!\over(2m_+-1)(2^{m_+} m_+!)^2}.
\eqnoi
$$
There is a strong similarity with the time dependent counterpart,
$\pmt$.  The distribution among aggregates with the same total mass,
$m_+$, is given by the same combinatorial factor $\gm$ which has
already appeared in solution to the time dependent problem.
For large masses, the distribution describing the sum variable,
$m_+$, is an algebraic one, $N^{-1}P(m_+)\sim m_+^{-3/2}$.
We can write now the asymptotic form of the complete steady
state distribution,
$$
P(\bm)\equiv P(m_+,m_-)\sim \sqrt{h}\,\,m_+^{-(k+2)/2} \exp(-m_-^2/2m_+).\eqnoi
$$
For the time dependent problem, an additional size scale emerges for the
mass difference variable, $m_-\sim\sqrt{m_+}$. In the steady state,
on the other hand, the
total mass diverges and thus, the central limit theorem does not apply.
However, the similarity between the two cases is still strong as for a
fixed $m_+$, the mass difference distribution function is Gaussian.

For completeness, we briefly discuss the relaxation properties of the
density towards the steady state. It is possible to obtain these
properties by incorporating the time dependent density, $N\sim
t^{-\a}$, and the steady state solution, $N\sim
h^{\g}$,  into a single expression. Let us assume the
scaling form [17]
$$
N(h,t)\sim t^{-\a}\psi(t/\tau),\eqnoi
$$
with the relaxation time $\tau\sim h^{-\b}$. In the limit of a
vanishing input rate the time dependent solution must be recovered and
thus, $\psi(x)\to 1$ as $x\to 0$. In the long time limit steady state
is approached and thus, $\psi(x)\sim x^{\a}$ for $x\gg1$.  Comparing
with the definition of the relaxation time the exponent relation
$$
\g=\a\b\eqnoi
$$
is found. The exponent $\b=1/2$, is obtained from the decay exponent,
$\a=1$, and the steady state exponent, $\g=1/2$.  Hence, the
relaxation time diverges in the limit of a vanishing input rate, $\tau
\sim h^{-1/2}$. This result can be obtained directly by solving the
equation $dN/dt=-N^2+h$ which is readily performed to find
\hbox{$N(h,t)=\sqrt{h}\tanh\big((t+t_0)\sqrt{h}\big)$}.
The time shift $t_0$ is determined by the initial density,
$N(t=0)=\sqrt{h}\tanh(\sqrt{h}t_0)$. We prefer the above scaling
argument since it is applicable to a wide class of problems.

\vfill\eject

\centerline{\bf III. Ballistic Aggregation}\medskip

Ballistic aggregation is a natural and important example of a process
with multiple conservation laws.  Both the mass and the momentum are
conserved quantities and thus, there are $d+1$ conservation laws in
$d$ dimension [26].  A collision between two particles results in an
aggregate whose mass as well as momentum are given by a sum over its
components. One can view this system as a gas of sticky particles,
\ie, an inelastic gas with a vanishing restitution
coefficient. Heuristic arguments predict  certain
scaling properties of the system.  However, other aspects of this
problem, such as the mass-momentum distribution function remain
unsolved. Our theory is well suited for this problem and yields
an approximate form for the particle distribution function.

We start by discussing the one dimensional case and then generalize to
higher dimensions. We denote the probability distribution function for
particles of mass $m$ and momentum $p$ at time $t$ by $\pmp$ (we
suppress the time variable).  The {\it stosszahlansatz}
Boltzmann equation, which
describes the temporal evolution of this density, must conserve mass
and momentum. Additionally, the collision rate between two particles
is given by their velocity difference and hence, the Boltzmann
equation reads
$$
\td {\pmp} t =
\sum_{m',p'}\left|v'-v''\right|P(m',p')P(m-m',p-p')
-2\pmp\sum_{m',p'} \left|v-v'\right|P(m',p'),\eqnoi
$$
where $v'-v''=p'/m'-(p-p')/(m-m')$ and $v-v'=p/m-p'/m'$. This
approximation ignores possible spatial correlations and is generally
uncontrolled.  By replacing the kernel terms $|v-v'|$ and $|v'-v''|$
with their average value $\vav$, an approximate Boltzmann equation
is written,
$$
{d\pmp\over dt}=
\vav\left(\sum P(m',p')P(m-m',p-p')- 2 \pmp \sum P(m',p')\right). \eqnoi
$$
The time-dependent factor $\vav$ can be absorbed into a novel time
variable, $T$, defined by $dT/dt=\vav$. The resulting Boltzmann
equation is equivalent to Eq.~(1).

For simplicity we consider an initial system of identical particles
with zero average momentum. Without loss of generality we assume that
typical initial quantities such as the mass and the momentum equal
unity. Hence, the initial distribution function, $P_0(m,p)$, is given
by
$$
P_0(m,p)=\delta(m-1)\big(\delta(p-1)+\delta(p+1)\big)/2.\eqnoi
$$
In principle, the general solution of \back1\ can be obtained using
the generating function method. However,
with these specific initial conditions, the process reduces to the two-mass
process discussed in the preceding section. Indeed, by identifying $m$
with the total mass $m_+=m_1+m_2$ and $p$ with the mass difference
$m_-=m_1-m_2$, the initial conditions Eq.~(6) and \last\ are the same.
Moreover, identical rate equations describe the time
evolution of both $P(m_1,m_2)$ and $\pmp$. From the solution of
Eq.~(11), the mass-momentum distribution function is found in terms of the
time variable $T$,
$$
P(p,m)={T^{m-1}\over (1+T)^{m+1}}\,
\,{2^{-m}m!\over \left((m+p)/2\right)!\left((m-p)/2\right)!}.
\eqnoi
$$

Using the asymptotic scaling properties of the solution $m\sim T$ and
$p\sim\sqrt{T}$, we can rewrite the solution in terms of the observable time
$t$. Since $dT/dt\sim v\sim p/m\sim T^{-1/2}$, one has $T\sim t^{2/3}$.
Hence, the well known the scaling laws
$$
m\sim t^{2/3}\qquad{\rm and}\qquad p\sim t^{1/3},\eqnoi
$$
are recovered. Asymptotically, the mass-momentum distribution is given
by the following form
$$
P(p,m)\sim
\mav^{-2}m^{-1/2}\exp\Bigl(-{m\over \mav}-{p^2\over 2m}\Bigr),
\eqnoi
$$
with the average mass $\mav\sim t^{2/3}$.
As discussed in the previous section, the momentum $p$ is a sum of $m$
independent variables. As a result, for a fixed mass $m$
the momentum distribution is Gaussian and $p\sim\sqrt{m}$. From
Eq.~(10), the mass distribution is [27]
$P(m)=\mav^{-2}\exp\big(-m/\mav\big)$.
Direct integration of \last\ shows that the momentum distribution is
also purely exponential, $P(p)=\big(\mav\pav\big)^{-1}\exp(-|p|/\pav)$, with
the typical momentum, $\pav=\sqrt{\mav/2}$.
On the other hand, the velocity distribution is
algebraic for large velocities $p(v)\sim |v|^{-3}$.

It is interesting to compare these predictions with numerical
simulations.  Carnevale \ea\ [26] established a scaling behavior
of \back1\ heuristically and confirmed it numerically.
They also reported a distribution that is reminiscent of \last.
Their numeric form resembles \last\ in that the mass distribution
is exponential and in that for a fixed mass the velocity
distribution is Gaussian. However, there is a significant
difference between the two forms as the simulation data suggests that
the velocity distribution is  independent of mass. This observation is an
intriguing one, since for a fixed mass $m$ the typical velocity is
mass dependent $v(m)\sim m^{-1/2}$. Jiang and Leyvraz [28] also studied
the mass distribution and found that it is singular near the origin,
$P(m)\sim m^{-1/2}$ for $m\ll\mav$, in contradiction with our
approximate theory. Curiously, the mass-momentum distribution contains
an identical singularity. However, this singularity disappears when the
momentum is integrated out.

A similar line of reasoning applies in $d$ dimensions.  We assume that
the density of an aggregate is constant, and thus, as the aggregation
process evolves, the size of an aggregate grows
indefinitely. Initially, only monomers with unit momentum occupy the
system.  Since the collision rate is also proportional  to the surface
area of an aggregate, the typical  collision rate is $va^{d-1}$,
with the radius $a\sim m^{1/d}$. The density satisfies the approximate
Boltzmann equation
$$
\td {P(m,\bp)} t=\langle va^{d-1}\rangle
\left(\sum P(m',\bp')P(m-m',\bp-\bp')-
2 P(m,\bp) \sum P(m',\bp')\right), \eqnoi
$$
where $\bp$ is the $d$-dimensional momentum.  Repeating the above
analysis yield the leading scaling behavior for the mass and the
momentum,
$$
m\sim t^{2d/(d+2)}\qquad{\rm and}\qquad |\bp|\sim t^{d/(d+2)}.\eqnoi
$$
The distribution function is a simple generalization of \back1,
$$
P(m,\bp)\sim
\mav^{-2}m^{-d/2}\exp\Bigl(-{m\over \mav}-{d|\bp|^2\over 2m}\Bigr),
\eqnoi
$$
with the average mass $\mav\sim t^{2d/(d+2)}$.
\back1 suggests that ballistic aggregation has no upper critical dimension.
Hence, it is not clear whether our approximation holds in sufficiently
large dimension, as is the case for diffusive aggregation.  In a
recent study of the two-dimensional gas of sticky particles [30], some
deviations from the mean-field predictions have been observed. The
simulations [30] revealed that the growth exponent varies with the initial
density. However, an exponential mass distribution and a Boltzmann
energy distribution were found, in agreement with mean-field
theory. We conclude that despite the crude nature of the
approximation, it provides good estimates for the leading asymptotic
behavior as well as the various distribution functions.

Steady state can be achieved by adding particles to the system with
rate $h$. We consider homogeneous and isotropic input of particles
with unit mass and unit momentum.  From Eq.~(21), the
distribution function reads,
$$
P(m,p)=\sqrt{h}\,\,m^{-(d+3)/2}
\exp\Bigl(-{d|\bp|^2\over 2m}\Bigr).
\eqnoi
$$
Furthermore, the mass distribution is given by $N^{-1}P(m)\sim
m^{-3/2}$, with the density $N\sim \sqrt{h}$. Note that the velocity
kernel $\vav$ is not important in the steady state since $\vav\propto \sum
P(m)v(m)\sim \int m^{-3/2}m^{-1/2}$ is finite. One can also study the
relaxation properties of the system. The relaxation towards the
steady state becomes slower and slower as the input rate vanishes.
Following the analysis of Eq.~(22), the corresponding
relaxation exponent is obtained,
$$
\tau(h)\sim h^{-\b},\qquad{\rm with}\qquad \b={d+2\over 4d}.
\eqnoi
$$
However, only in the low input rate limit, $h\ll 1$, steady state
is achieved. Indeed, the previous description is valid in the low
coverage limit, \ie, when $t\ll t_c$ with $t_c\sim h^{-1}$,
while for $t>t_c$ the space is covered by a single ``superparticle''.
Since the relaxation exponent $\b$ is less than one, the time scales
$\tau$ and $t_c$ are well separated in the low input rate limit.
Thus, the steady state distribution of \back1\ provides
an intermediate asymptotics valid for $\tau\ll t\ll t_c$.

In contrast, the nature of steady state caused by a spatially {\it
localized} particle input is a truly asymptotic one. The spatial
dependence of the density distribution is of particular interest in
this problem.  We consider a spherically symmetric point-like source
of ballistic particles with unit mass and unit momentum.  Let
$P(m,\bp,r)=P(m,\bp,r,t=\infty)$ be the steady-state radial
mass-momentum density.  For $d>2$ the reaction is in fact irrelevant
away from the source as will become clear below.
The density satisfies the convection equation,
$d\bigl(r^{d-1}P(m,\bp,r)\bigr)/dr=0$, and thus, the
concentration decays as $r^{-(d-1)}$. So in the {\it inhomogeneous}
ballistic problem, $d=2$ is a critical dimension above which the
reaction merely leads to the renormalization of the strength of the
source.

For $d\le2$, away from the source, particles have typically collided
many times and they move with a velocity close to unity in the radial
direction.  Thus, collisions occur with rate proportional to the rms
velocity $v$. The steady-state radial distribution $P(m,\bp,r)$
satisfies
$$
{1\over r^{d-1}}{d\over dr} \bigl(r^{d-1}P(m,\bp)\bigr)=\langle
va^{d-1}\rangle
\left(\sum P(m',\bp')P(m-m',\bp-\bp')-
2 P(m,\bp) \sum P(m',\bp')\right).
\eqnoi
$$
It is not difficult to verify that with the transformation
$r^{d-1}P(m,\bp)\to P(m,\bp)$ and $R=\int^r r'^{(1-d)}dr'\to t$,
the steady-state equation reduces to the time-dependent
equation (31). Therefore when $d<2$, the variable $R\sim r^{2-d}$
plays the role of time and from Eq.~(32),
$m(r)\sim R^{2d/(2+d)}\sim r^{2d(2-d)/(2+d)}$.
As $d\to 2$, the exponent describing the mass growth vanishes,
indicating that $d=2$ is indeed the critical dimension,
above which the typical mass far from the source is constant.
At the critical dimension $d=2$, logarithmic behavior occurs,
$R\sim\log(r)$, and consequently, $m(r)\sim R \sim \log(r)$.
Hence, for $d\le 2$ clustering is significant and the typical
mass is a growing function of the distance from the source.

To determine the mass-momentum distribution, we tacitly impose
boundary conditions similar to the initial conditions of the time
dependent problem, $r_0^{d-1}P(m,\bp,r_0)=\delta(m-1)\delta(|\bp|-1)$.
The steady-state density far from the source reads
$$
P(m,\bp,r)\sim r^{-(d-1)}\mavr^{-2} m^{-d/2}
\exp\left(-{m \over \mavr}-{d(|\bp|-m)^2\over 2m}\right)\qquad d\le 2,
\eqnoi.
$$
with $\mavr\sim r^{2d(2-d)/(2+d)}$ for $d<2$ and $\mavr\sim\log(r)$
for $d=2$.  For a fixed mass, the momentum distribution is Gaussian
and as a result, the rms momentum is characterized by
$\sqrt{m(r)}$. By integration of the mass-momentum density, one finds
the concentration, $N(r)\sim 1/(r^{(d-1)}\mavr)$.  As a result,
$N(r)\sim r^{(2-5d+d^2)/(2+d)}$ for $d<2$, and $N(r)\sim
1/\big(r\log(r)\big)$ for $d=2$.

Returning to the time dependent problem, the steady state solution
holds for $r<t$ only, while for a larger
the space is essentially empty. It is useful to estimate the total
number of clusters $\cnt\sim \int_0^t r^{d-1}N(r)$.
For $d<2$ one finds $\cnt\sim t^{(2d^2-3d+2)/(d+2)}$, and for
$d=2$ one has $\cnt\sim t/\log(t)$. In the ballistic regime, $d>2$,
collisions do not cause a significant reduction in the number of clusters,
and the total number of clusters grows linearly in time.

To summarize, we write the leading asymptotic behaviors of the mass
$$
m(r)\sim\cases{r^{2d(2-d)/(2+d)}&$d<2$;\cr
               \log(r) &$d=2$;\cr
               1  &$d>2$,\cr}\eqnoi
$$
the  density
$$
N(r)\sim\cases{r^{-(5d-d^2-2)/(2+d)}&$d<2$;\cr
               r^{-1}[\log(r)]^{-1} &$d=2$;\cr
                r^{-(d-1)}  &$d>2$,\cr}\eqnoi
$$
and the total number of clusters
$$
\cnt \sim\cases{t^{(2d^2-3d+2)/(2+d)}&$d<2$;\cr
                   t/\log(t) &$d=2$;\cr
                   t  &$d>2$.\cr}\eqnoi
$$
The typical rms momentum behaves as $\sqrt{m(r)}$. Numerical simulations
agree with the above in $d=1$ [26]. It will be interesting to test
these predictions in higher dimensions.

\vfill\eject
\centerline{\bf IV. Diffusive Aggregation}\medskip

In this section, we consider the diffusive driven
aggregation processes involving $k$ distinct species.
Each of the species masses is conserved and thus there
are $k$ conservation laws.
We apply the rate theory described in Sec.~II and also
investigate low dimensional systems where the rate equation
description is expected to fail [15]. In all dimensions we
find that in addition to the typical mass
scale, there exists an additional mass scale.

In spatial dimensions larger than the critical dimension, $d_c=2$,
Smoluchowski rate theory is exact.  In dimensions lower than the
critical dimension, spatial correlations are significant
asymptotically.  Particles are repelling each other, and the
effective reaction rate, $\k$, depends on the density [20]
$$
\k\sim\cases{ N^{2/d-1}&$d<2$;\cr
                1/|\log(N)| &$d=2$;\cr
                1     &$d>2$.\cr}\eqnoi
$$
This reaction rate can be absorbed into a suitably defined time
variable $T$,
$$
\td T t = \k . \eqnoi
$$
With this novel time, Smoluchowski rate equation reduces to Eq.~(1).

For $d\le d_c$, this approximation yields erroneous results for the mass
distribution. Nevertheless, it produces correct asymptotic scaling
behavior for quantities such as the typical mass. Thus for the general
case $d\le d_c$ we just quote the leading asymptotic
behaviors while in one dimension we provide exact results.
Let us consider the case where there are  two species,
say $A$ and $B$.  A cluster is characterized by the respective
masses of its components, $m_A$ and $m_B$.  The typical total mass
is given by $m_A+m_B\sim T$, or equivalently,
$$
m_A+m_B\sim\cases{ t^{d/2}&$d<2$;\cr
                   t/\log(t) &$d=2$;\cr
                   t  &$d>2$.\cr}\eqnoi
$$
On the other hand, since the mass difference is a Gaussian variable,
for a fixed total mass, one has $|m_A-m_B|\sim\sqrt{m_A+m_B}$. This
additional mass scale can be also expressed in terms of time,
$$
|m_A-m_B|\sim\cases{ t^{d/4}&$d<2$;\cr
                   \sqrt{t/\log(t)} &$d=2$;\cr
                   \sqrt{t}  &$d>2$.\cr}\eqnoi
$$
Note also that the present two-mass aggregation process can be mapped
onto a two-species aggregation-annihilation process with two
conservation laws [23].  Indeed, the above results agree with
analytical and numerical findings of Refs.~[23,32].

In one dimension, it is possible to derive a complete analytical
solution. We consider a linear lattice on which point clusters
hop randomly from site to nearest neighbor sites. We assume that
the diffusion coefficient $D$ does not depend on cluster's mass.
We also assume that initially each site is occupied by some monomer,
$A$ or $B$ with equal probability. The multivariate distribution
function can be expressed in the form of Eq.~(11), with the sum
variable given by the Spouge's solution [19] of the ordinary
one dimensional diffusion-controlled aggregation problem,
$$
P(m_+,t)=e^{-4Dt}\bigl[I_{m_+-1}(4Dt)-I_{m_++1}(4Dt)\bigr],
\eqnoi
$$
where $I_n$ denotes a modified Bessel function. Introducing now the
scaling variables, $M_+=m_+/\sqrt{8Dt}$ and $M_-=m_-/(8Dt)^{1/4}$,
one can rewrite the solution in a convenient scaling form
$$
P(m_1,m_2,t)\sim t^{-5/4} \Phi(M_+,M_-),\eqnoi
$$
with the scaling function $\Phi$
$$
\Phi(x,y)=\sqrt{x}\exp\Bigl(-x^2-{y^2\over 2x}\Bigr).
\eqnoi
$$

Finally, we consider diffusive aggregation
in a system with a steady spatially localized monomer input.
We assume that monomers of all types are added at random
with an equal rate. Making use of exact and scaling
results for the corresponding ordinary aggregation [20], we
solve for our case. This inhomogeneous system is
characterized by two critical dimensions, the usual ``homogeneous''
critical dimension $d_c=2$ and an additional critical dimension
$d^c=4$ which demarcates the pure diffusion regime $d>4$
(particles do not affect each other far away from the source)
and the diffusion-reaction regime $d\le 4$. The system approaches
steady state as $t\to \infty$. In the diffusion-reaction
regime, the density distribution function approaches a power-law
in $r$, where $r$ is the distance from the source. The borderline
cases $d=d_c=2$ and $d=d^c=4$ should be treated more carefully
since logarithmic factors appear. We write the final results
for the typical total mass
$$
m_A+m_B\sim\cases{ r^2&$d<2$;\cr
                   r^2/\log(r) &$d=2$;\cr
                   r^{4-d}  &$2<d<4$;\cr
                   \log(r) &$d=4$;\cr
                   1       &$d>4$.\cr}\eqnoi
$$
For a fixed total mass $m_A+m_B$, the mass difference is again Gaussian
and consequently, $|m_A-m_B|\sim\sqrt{m_A+m_B}$.

\vfill\eject
\bigskip\centerline{\bf V. Conclusions}\medskip

In summary, we have investigated irreversible aggregation with many
conservation laws.  The solution to this process is characterized by
the Gaussian statistics of the fluctuations in a given conserved
quantity. The process is thus governed asymptotically by two size
scales. Application to a ``sticky gas'' suggests a Boltzmann velocity
distribution for a fixed mass. In addition, the mass distribution is
exponential. By comparing our predictions with available numerical
results we have found that the present approximate theory gives a good
description of the sticky gas. We have also investigated steady-state
properties by introducing a localized source. We have observed two
different behaviors, the ballistic-reaction regime for $d\le 2$, and
the pure ballistic regime for $d>2$. Application to diffusive
aggregation with more than one species also exhibit a novel
``diffusive'' size scale.

This study suggests different avenues for further investigation.
The theory might be applicable to problems such as catalysis and
chemical reactions with many species. It will be also interesting
to analyze the rate equations with more realistic reaction rates.
An important question to be addressed is how robust is the Gaussian
nature of the fluctuations statistics. It is plausible that spatial
correlations introduce nontrivial internal arrangement of the
clusters, leading to more complicated statistics. It is also
plausible that even on the rate equation level but with the
reaction rate not expressible as a function of the sum-variables
only, different asymptotic behavior emerges. This could
explain why for a dual fragmentation process the multivariate
generalization produces an infinite set of scales [33-35] compared
to the two scales in the models of multivariate aggregation
we have examined in this study.

\bigskip\centerline{\bf Acknowledgments}\medskip

We thank Y.~Du, L.~P.~Kadanoff, S.~Redner, and W.~R.~Young for useful
discussions.  P.L.K. was supported by ARO grant \#DAAH04-93-G-0021
and NSF grant \#DMR-9219845.  E.B. was supported in part by the MRSEC
Program of the National Science Foundation under Award Number
DMR-9400379 and by NSF under Award Number 92-08527.

\vfill\eject
\centerline{\bf References}

\refi P.~G.~de Gennes, {\sl Scaling Concepts in Polymer Physics}
      (Cornell Univ Press, Ithaca, 1979).
\refi F.~Family and D.~P.~Landau, {\sl Kinetics of Aggregation and Gelation}
      (North-Holland, New York, 1984).
\refi T.~A.~Witten and L.~Sander, \prl 47 1400 1981 .
\refi M.~Y.~Lin, H.~N.~Lindsay, D.~A.~Weitz, R.~C.~Ball, and R.~Klein,
      \nat 339 94 1987 .
\refi S.~K.~Frielander, {\sl Smoke, Dust  and Haze: Fundamentals of Aerosol
      Behavior} (Wiley, New York, 1977).
\refi M.~V.~Smoluchowski, \zpc 92 215 1917 .
\refi S.~Chandrasekhar, \rmp 15 1 1943 .
\refi F.~Leyvraz and H.~R.~Tschudi, \jpa 14 3389 1981 .
\refi M.~H.~Ernst and P.~G.~van Dongen, \prl 54 1396 1985 .
\refi D.~Tousaint and F.~Wilszek, \jcp 78 2642 1983 .
\refi D.~C.~Torney and H.~M.~McConnel, \prslA 387 147 1983 .
\refi P.~Meakin, \prl 51 1119 1983 .
\refi T.~Vicsek and F.~Family, \prl 52 1669 1984 .
\refi K.~Kang and S.~Redner, \prl 52 955 1984 .
\refi K.~Kang and S.~Redner, \pra 32 435 1985 .
\refi F.~Leyvraz and S.~Redner, \pra 46 3132 1992 .
\refi Z.~R\'acz, \prl 55 1707 1985 .
\refi A.~A.~Lushnikov, \jetp 64 811 1986 .
\refi J.~L.~Spouge, \prl 60 871 1988 .
\refi Z.~Cheng, S.~Redner, and F.~Leyvraz, \prl 62 2321 1989 .
\refi D.~ben-Avraham M.~A.~Burschka, and C.~R.~Doering, \jsp 60 695 1990 .
\refi J.~G.~Amar and F. Family, \pra 41 3258 1990  .
\refi P.~L.~Krapivsky, \pA 198 135 1993 ; \pA 198 150 1993 ;
      \pA 198 157 1993 .
\refi S.~F.~Shandarin and Ya.~B.~Zeldovich, \rmp 61 185 1989 .
\refi E.~Ben-Naim, P.~L.~Krapivsky and S.~Redner, \pre 50 822 1994 .
\refi G.~F.~Carnevale, Y.~Pomeau and W.~R.~Young, \prl 64 2913 1990 .
\refi J.~Piasecki, \pA 190 95 1992 .
\refi Y.~Jiang and F.~Leyvraz, \jpa 26 L179 1993 .
\refi P.~A.~Philippe and J.~Piasecki, \jsp 76 447 1994 .
\refi E.~Trizac and J.-P.~Hansen, \prl 74 4114 1995 .
\refi R.~P.~Treat, \jpa 23 3003 1990 .
\refi I.~M.~Sokolov and A.~Blumen, \pre 50 2335 1994 .
\refi P.~L.~Krapivsky and E.~Ben-Naim, \pre 50 3502 1994 .
\refi G.~J.~Rodgers and M.~K.~Hasan, \pre 50 3458 1994 .
\refi D.~Boyer, G.~Tarjus, and P.~Viot, \pre 51 1043 1995 .

\end